\documentclass[twocolumn, tighten]{aastex631}
\usepackage{float}
\usepackage[T1]{fontenc} % Ensures proper encoding
\usepackage[utf8]{inputenc} % Enables UTF-8 encoding

\usepackage{nameref}
\usepackage{comment}
\usepackage{booktabs}
\usepackage{cuted}
\usepackage{placeins}
\usepackage{hyperref}
\usepackage{amsmath}
\usepackage{float}
\usepackage{graphicx}

\usepackage{longtable}
\def\d   {{d $\mkern-14.3mu \mathchar'26 $}}

\begin{document}
\title{Search for the Blazhko effect in field RR Lyrae stars using LINEAR and ZTF light curves}

\author{Ema Donev}
\affiliation{XV. Gymnasium (MIOC), Jordanovac 8, 10000, Zagreb, Croatia}

\author[0000-0001-5250-2633]{\v{Z}eljko Ivezi\'{c}}
\affiliation{Department of Astronomy and the DiRAC Institute, University of Washington, 3910 15th Avenue NE, Seattle, WA 98195, USA}

\correspondingauthor{\v{Z}eljko Ivezi\'{c}}
\email{ivezic@uw.edu}
% Ema Donev: emadonev@icloud.com

\begin{abstract}
We analyzed the incidence and properties of RR Lyrae stars that show evidence for amplitude and
phase modulation (the so-called Blazhko Effect) in a sample of $\sim$3,000 stars with LINEAR and ZTF light
curve data collected during the periods of 2002-2008 and 2018-2023, respectively. A preliminary subsample
of about $\sim$500 stars was algorithmically pre-selected using various data quality and light curve statistics,
and then 228 stars were confirmed visually as displaying the Blazhko effect. This sample increases the number
of field RR Lyrae stars displaying the Blazhko effect by more than 50\% and places a lower
limit of (11.4$\pm$0.8)\% for their incidence rate. We find that ab type RR Lyrae that show the Blazhko effect have
about 5\% (0.030 day) shorter periods than starting sample, a 7.1$\sigma$ statistically significant
difference. We find no significant differences in their light curve amplitudes and apparent magnitude (essentially,
signal-to-noise ratio) distributions. No period or other differences are found for c type RR Lyrae. We find convincing
examples of stars where the Blazhko effect can appear and disappear on time scales of several years.  With
time-resolved photometry expected from LSST, a similar analysis will be performed for even larger samples of field
RR Lyrae stars in the southern sky and we anticipate a higher fraction of discovered Blazhko stars due to better
sampling and superior photometric quality.
\end{abstract}
\keywords{Variable stars --- RR Lyrae variable stars --- Blazhko effect}

\section{Introduction\label{sec:intro}}

RR Lyrae stars are pulsating variable stars with periods in the range of 3--30 hours and large amplitudes
that increase towards blue optical bands (e.g., in the SDSS $g$ band from 0.2 mag to 1.5 mag;
\citealt{2010ApJ...708..717S}). For comprehensive reviews of RR Lyrae stars, we refer the reader to \cite{1995CAS....27.....S} and \cite{2009Ap&SS.320..261C}.

RR Lyrae stars often exhibit amplitude and phase modulation, or the so-called Blazhko effect\footnote{The Blazhko effect was
probably discovered by Lidiya Petrovna Tseraskaya and first reported by Sergey Blazhko though exact discovery details remain unclear.} (hereafter,
``Blazhko stars''). For examples of well-sampled observed light curves showing the Blazhko effect, see,  e.g., Kepler
data shown in Figures 1 and 2 from \cite{2010MNRAS.409.1585B}. The Blazhko effect has been known for a long time \citep{1907AN....175..325B}, but its detailed observational properties and theoretical explanation of its causes remain elusive
\citep{2008JPhCS.118a2060K,2009AIPC.1170..261K,2014IAUS..301..241S}.
Various proposed models for the Blazhko effect, and principal reasons why they fail to explain observations, are summarized in \cite{2016CoKon.105...61K}. 

A part of the reason for the incomplete observational description of the Blazhko effect is difficulties in discovering a large number 
of Blazhko stars due to temporal baselines that are too short and insufficient number of observations per object
\citep{2016CoKon.105...61K,2022ApJS..258....4H}. With the advent of modern sky surveys, several studies
reported large increases in the number of known Blazhko stars, starting with a sample of about 700 Blazhko
stars discovered by the MACHO survey towards the LMC \citep{2003ApJ...598..597A} and about 500 Blazhko stars
discovered by the OGLE-II survey towards the Galactic bulge \citep{2003AcA....53..307M}. 
Most recently,  about 4,000 Blazhko stars were discovered in the LMC and SMC 
\citep{2009AcA....59....1S, 2010AcA....60..165S}, and an additional $\sim$3,500 stars were discovered in the
Galactic bulge \citep{2011AcA....61....1S, 2017MNRAS.466.2602P}, both by the OGLE-III survey. Nevertheless, discovering the Blazhko
effect in field RR Lyrae stars that are spread over the entire sky remains a much harder problem: only about
400 Blazhko stars in total \citep{2013A&A...549A.101S} from all the studies of field RR Lyrae stars have been reported so far (see also Table 1
in \citealt{2016CoKon.105...61K}). 

Here, we report the results of a search for the Blazhko effect in a sample of $\sim$3,000 field RR Lyrae stars with
LINEAR and ZTF light curve data. A preliminary subsample of about $\sim$500 stars was selected using various
light curve statistics, and then 228 stars were confirmed visually as displaying the Blazhko effect. This new
sample doubles the number of field RR Lyrae stars that exhibit the Blazhko effect. In \S\ref{sec:data}
and \S\ref{sec:analysis} we describe our datasets and analysis methodology, and in \S\ref{sec:results} we present our analysis results. 
Our main results are summarized and discussed in \S\ref{sec:discussion}.

\phantom{There is some latex bug somewhere and this dummy call is needed to force it to make pdf...}

\begin{figure*}[ht]
  \centering
  \includegraphics[width=18cm]{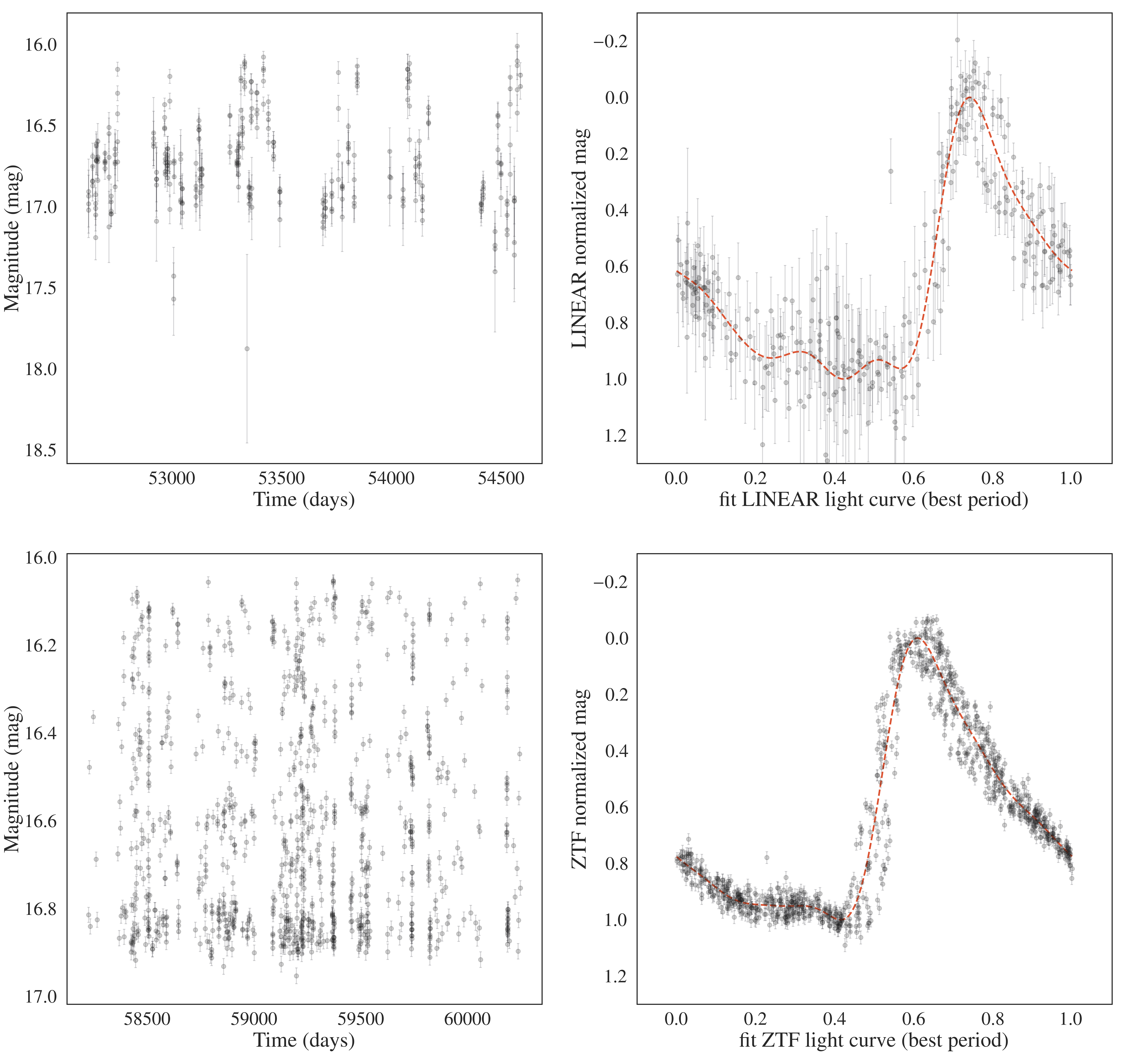}
  \vskip -0.2in
  \caption{An example of a Blazhko star (LINEARid = 136668) with
    LINEAR (top row; period = 0.532923 day) and ZTF (bottom row; period = 0.532929 day) light
    curves (left panels, data points with ``error bars''), phased light curves normalized to the 0--1 range (right panels, data points
    with ``error bars''), with their best-fit models shown by dashed lines. The best-fit period is determined for each
    dataset separately using 3 Fourier terms. The models shown in the right panels are evaluated with 6 Fourier terms. }
 \label{fig:lc_pair}
\end{figure*}
 
\section{Data Description and Period Estimation \label{sec:data}}

Analysis of field RR Lyrae stars requires a sensitive time-domain photometric survey over a large sky area.
For our starting sample, we used $\sim$3,000 field RR Lyrae stars with light curves obtained by the LINEAR
asteroid survey. In order to study long-term changes in light curves, we also utilized light curves obtained
by the ZTF survey which monitored the sky $\sim$15 years after LINEAR. The combination of LINEAR and
ZTF provided a unique opportunity to systematically search for the Blazhko effect in a large number of
field RR Lyrae stars over a large time span of two decades. 

We first describe each dataset in more detail, and then introduce our analysis methods. All our analysis
code, written in Python, is available on GitHub\footnote{\url{https://github.com/emadonev/var_stars}}.  
 
\subsection{LINEAR Dataset}

The properties of the LINEAR asteroid survey and its photometric re-calibration based on SDSS data are discussed
in \cite{2011AJ....142..190S}. Briefly, the LINEAR survey covered about 10,000 deg$^2$ of the northern sky in white
light (no filters were used, see Fig.~1 in \citealt{2011AJ....142..190S}), with photometric errors ranging from $\sim$0.03
mag at an equivalent SDSS magnitude of $r=15$ to 0.20 mag at $r\sim18$. Light curves used in this work include,
on average, 270 data points collected between December 2002 and September 2008.
 
A sample of 7,010 periodic variable stars with $r<17$ discovered in LINEAR data were robustly classified by
\cite{2013AJ....146..101P}, including
about $\sim$3,000 field RR Lyrae stars of both ab and c type, detected to distances of about 30 kpc \citep{2013AJ....146...21S}.
The sample used in this work contains 2196 ab-type and 745 c-type RR Lyrae, selected using classification labels and the {\it gi}
color index from \cite{2013AJ....146..101P}.
The LINEAR light curves, augmented with IDs, equatorial coordinates, and other data, were accessed using the astroML Python
module\footnote{For an example of light curves, see \url{https://www.astroml.org/book_figures/chapter10/fig_LINEAR_LS.html}}
\citep{2012cidu.conf...47V}.

\subsection{ZTF Dataset}

The Zwicky Transient Factory (ZTF) is an optical time-domain survey that uses the Palomar 48-inch Schmidt telescope
and a camera with 47 deg$^2$ field of view \citep{2019PASP..131a8002B}. The dataset analyzed here was obtained with
SDSS-like $g$, $r$, and $i$ band filters. Light curves for objects in common with the LINEAR RR Lyrae sample typically
have smaller random photometric errors than LINEAR light curves because ZTF data are deeper (compared to LINEAR,
ZTF data have about 2-3 magnitudes fainter  $5\sigma$ depth). ZTF data used in this work were collected between
February 2018 and December 2023, on average about 15 years after obtaining LINEAR data. The median number of
 observations per star for ZTF light curves is $\sim$500. 

The ZTF dataset for this project was created by selecting ZTF IDs with matching equatorial coordinates to a corresponding
LINEAR ID of an RR Lyrae star. This process used the {\it ztfquery} function, which searched the coordinates in the ZTF database
within 3 arcsec from the LINEAR position. The resulting sample consisted of 2857 RR Lyrae stars with both LINEAR and ZTF data.
The fractions of RRab and RRc type RR Lyrae in this sample, 71\% RRab and 29\% RRc type, are consistent with results from
other surveys \citep[e.g.,][]{2010ApJ...708..717S, 2023A&A...674A..18C}.

\subsection{Period Estimation}

The first step of our analysis is estimating best-fit periods, separately for LINEAR and ZTF datasets. 
We used the Lomb-Scargle method \citep{2015zndo.....14833V} as implemented in {\it astropy}
\citep{2018AJ....156..123A}. The period estimation used 3 Fourier components and a two-step process: an initial
best-fit frequency was determined using the {\it autopower} frequency grid option and then the power spectrum was
recomputed around the initial frequency using an order of magnitude smaller frequency step. In case of ZTF, we
estimated period separately for each available passband and adopted their median value. Once the best-fit
period was determined, a best-fit model for the phased light curve was computed using 6 Fourier components.
Fig \ref{fig:lc_pair} shows an example of a star with LINEAR and ZTF light curves, phased light curves, and their
best-fit models.  

We found excellent agreement between the best-fit periods estimated separately from LINEAR and ZTF light curves. 
The median of their ratio is unity within $2\times10^{-6}$ and the robust standard deviation of their ratio is
$2\times10^{-5}$. With a median sample period of 0.56 days, the implied scatter of period difference is about 1.0 sec.  

Given on average about 15 years between LINEAR and ZTF data sets, and a typical period of 0.56 days, this time
difference corresponds to about 10,000 oscillations. With a fractional period uncertainty of $2\times10^{-5}$,
LINEAR data can predict the phase of ZTF light curve with an uncertainty of 0.2. RR Lyrae light curves may experience
phase changes of this magnitude \citep[see e.g.,][]{2011MNRAS.411.1744S, 2017IBVS.6228....1D};  therefore, each
data set must be analyzed separately. On the other hand, amplitude modulation can be detected on time scales as
long as 15 years, as discussed in the following section. 

We did not try to identify double-mode (RRd) stars because their expected sample fraction is below 1\%
\citep[][]{2023A&A...674A..18C}. 

\section{Analysis Methodology: Searching for the Blazhko Effect  \label{sec:analysis}}  

\begin{figure}[ht]
\resizebox{\hsize}{!}{\includegraphics[width=17cm]{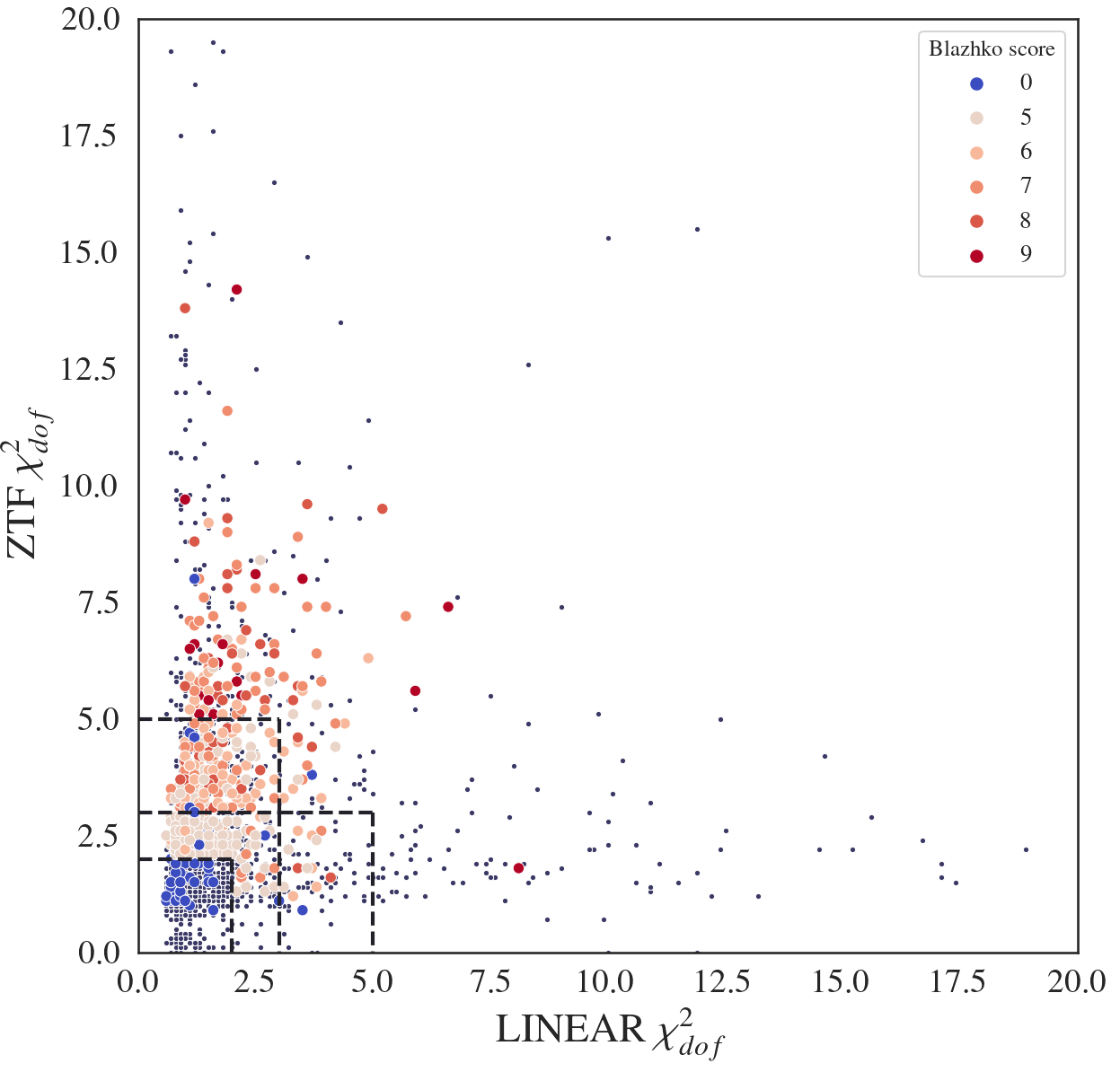}}
\caption{A selection diagram constructed with the two sets of robust $\chi^2_{dof}$ values, for LINEAR and ZTF data sets, where
 the dark blue dots represent all RR Lyrae stars and the circles represent candidate Blazhko stars (color-coded according to the
 legend, with B\_score representing the number of points scored from the selection algorithm).  The horizontal
  and vertical dashed lines help visualize selection boundaries for Blazhko candidates (see text).}
\label{fig:chi2}
\end{figure}

Given the two sets of light curves from LINEAR and ZTF, we searched for amplitude and phase modulation,
either during the 5-6 years of data taking by each survey, or during the average span of 15 years between the two
surveys. Starting with a sample of 2857 RR Lyrae stars, we pre-selected a smaller sample that was inspected
visually (see below for details). We also required at least 150 LINEAR data points and 150 ZTF data points (for the
selected band from which we calculated the period) in analyzed light curves. We used two pre-selection methods
that are sensitive to different types of light curve modulation: direct light curve analysis and periodogram analysis,
as follows.

\subsection{Direct Light Curve Analysis}

\begin{figure*}[ht]
    \centering
    \resizebox{\hsize}{!}{\includegraphics[width=21cm]{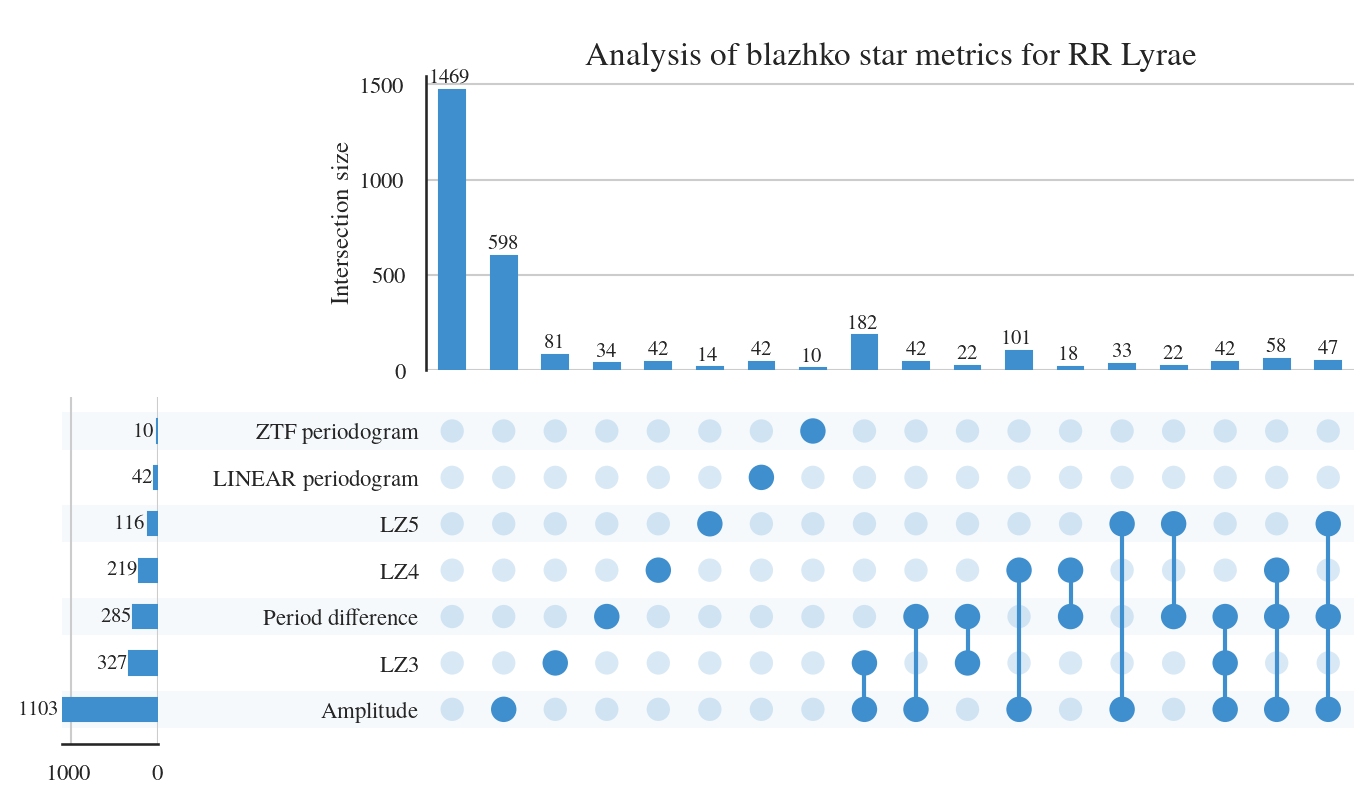}}
    \vskip -0.1in
    \caption{The figure shows selection criteria and the resulting numbers of pre-selected Blazhko star candidates for each
      criterion and their combinations (x in LZx corresponds to the number of scored points in the $\chi^2_{dof}$ vs. $\chi^2_{dof}$
      diagram (see Fig.~\ref{fig:chi2}). The dots represent each case a star can occupy, where every solid dot is a specific
      criterion that is satisfied. Connections between solid dots represent stars which satisfy multiple criteria. Each dot
      combination has its own count, represented by the horizontal countplot. The vertical countplot shows the total number
      of stars that satisfy one criteria (union of all cases). For example, a total of 116 stars passed the LZ5 criterion,  with 14
      of them satisfying only $\chi^2$ criterion, 33 also had a significant amplitude change, 22 had a significant period
      difference, and 47 had both a significant period and amplitude difference along with the satisfied $\chi^2$  criterion.
      The sum of all specific cases is 116.}
      \label{fig:selstats}
    \end{figure*}
    
Given statistically correct period, amplitude and light curve shape estimates,
as well as data being consistent with reported (presumably Gaussian) uncertainty estimates, the $\chi^2$ per degree
of freedom gives a quantitative assessment of the \textit{"goodness of fit"},
\begin{equation}
        \chi_{dof}^2 = {1 \over N_{dof}} \, \sum{\frac{(d_i - m_i)^2}{\sigma_i^2}}.
\end{equation}
Here, $d_i$ are measured light curve data values at times $t_i$, and with associated uncertainties $\sigma_i$,
$m_i$ are best-fit models at times $t_i$, and $N_{dof}$ is the number of degrees of freedom, essentially the
number of data points. In the absence of any light curve modulation, the expected value of $\chi^2_{dof}$ is
unity, with a standard deviation of $\sqrt{2/N_{dof}}$.  If $\chi^2_{dof} - 1$ is many times  larger than 
$\sqrt{2/N_{dof}}$, it is unlikely that data $d_i$ were generated by the assumed (unchanging) model $m_i$.  
Of course, $\chi^2_{dof}$ can also be large due to underestimated measurement uncertainties $\sigma_i$,
or to occasional non-Gaussian measurement error (the so-called outliers). 

Therefore, to search for signatures of the Blazhko effect, manifested through statistically unlikely large values
of $\chi^2_{dof}$, we computed $\chi^2_{dof}$ separately for LINEAR and ZTF data (see Fig.~\ref{fig:chi2}). 
Using the two sets of $\chi^2_{dof}$ values, we algorithmically pre-selected a sample of candidate Blazhko stars
for further visual analysis of their light curves. The visual analysis is needed to confirm the expected Blazhko behavior
in observed light curves, as well as to identify cases of data problems, such as photometric outliers. 

We used a simple scoring algorithm, optimized through trial and error, that utilized the two values of $\chi^2_{dof}$,
augmented by period and amplitude differences, as follows. A star could score a maximum of 9 points,
and a minimum of 5 points was required for further visual analysis. The $\chi^2_{dof}$ selection boundaries are 
illustrated in Fig.~\ref{fig:chi2}. If either value of $\chi^2_{dof}$
exceeded 5, or both exceeded 3, a star was awarded 5 points and immediately selected
for further analysis. If these $\chi^2_{dof}$ criteria were not met, a
star could still be selected by meeting less stringent $\chi^2_{dof}$
selection if it also had large period or amplitude difference between
LINEAR and ZTF datasets. Stars with at least one value of $\chi^2_{dof}$
above 2 would receive 3 points and those with at least one
$\chi^2_{dof}$ above 3 would receive 4 points. A period
difference exceeding $2\times10^{-4}$ day would be awarded 1 point
and two points for exceeding $5\times10^{-4}$ day. Analogous limits
for amplitude difference were 0.05 mag and 0.15 mag, respectively. 
 
The candidate Blazhko sample pre-selected using this method includes 531 stars. For most selected stars,
the $\chi^2_{dof}$ values were larger for the ZTF data because the ZTF photometric uncertainties are smaller
than for the LINEAR data set.  Fig.~\ref{fig:selstats} summarizes the selection criteria and the resulting numbers of
selected stars for each criterion and their combinations.

\subsection{Periodogram Analysis}

\begin{figure*}[ht]
  \centering
  \resizebox{\hsize}{!}{\includegraphics[width=14cm]{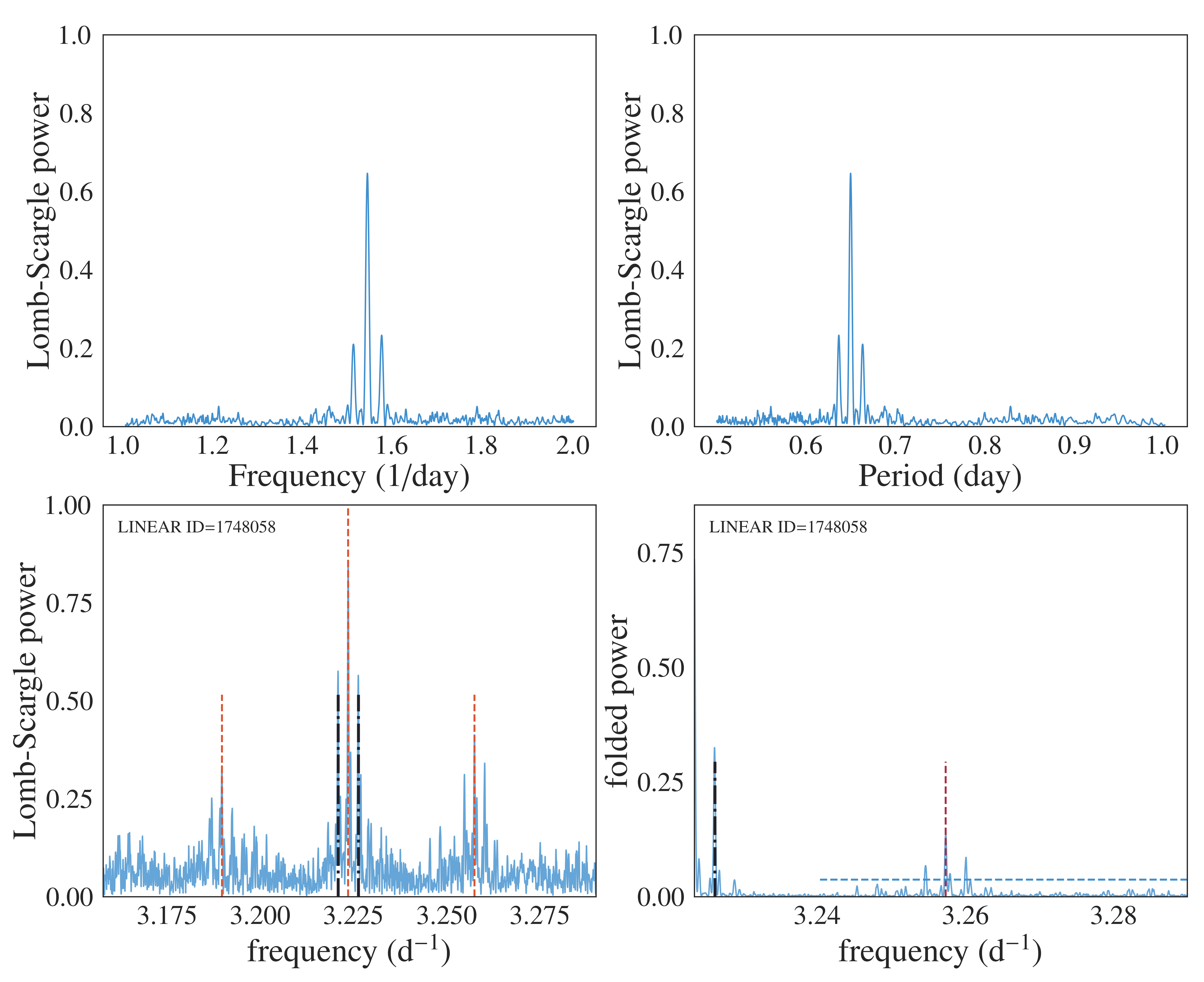}}
  \caption{The top two panels show a simulated periodogram for a sum of two {\it sine} functions with similar frequencies
    $f_1$ and  $f_2$ --   the central peak corresponds to their mean (see eqs.~\ref{eq:fo} and \ref{eq:Df}).
    The bottom left panel shows a periodogram for an observed LINEAR light curve for $ID=1748058$, and the bottom right panel shows its
    folded version (around the main frequency $f_o=3.223$ d$^{-1}$). In the bottom left panel, the three vertical dashed
    lines show the three  frequencies identified by the algorithm described in text, and the two dot-dashed lines mark
    yearly aliases around the main frequency $f_o$, at frequencies $f_o \pm 0.0274$ d$^{-1}$. The two vertical lines in the
    bottom right panel have the same meaning, and the horizontal dashed line shows the noise level multiplied by 5.}
\label{fig:periodogram}
\end{figure*}

When light curve modulation is due to double-mode oscillation with two similar oscillation frequencies (periods),
it is possible to recognize its signature in the periodogram computed as part of the Lomb-Scargle analysis. Depending
on various details, such as data sampling and the exact values of periods, amplitudes, this method may be
more efficient than direct light curve analysis \citep{2020MNRAS.494.1237S}. We also employed this method to select
additional candidates, as follows.

A sum of two {\it sine} functions with same amplitudes and with frequencies $f_1$ and $f_2$ can be rewritten 
using trigonometric equalities as 
\begin{equation}
         y(t) = 2 \, \cos(2\pi{f_1-f_2\over 2} t) \, \sin(2\pi {f_1+f_2\over 2} t).
\end{equation} 
We can define 
\begin{equation}
\label{eq:fo}
         f_o = {f_1+f_2\over 2},
\end{equation} 
and 
\begin{equation}
\label{eq:Df}
         \Delta f = |{f_1-f_2\over 2}|,
\end{equation} 
with $\Delta f << f_o$ when $f_1$ and $f_2$ are similar. The fact that $\Delta f$ is much smaller than $f_o$ means
that the period of the {\it cos} term
is much larger than the period of the basic oscillation ($f_o$). In other words, the {\it cos} term acts as a slow
amplitude modulation of the basic oscillation. When the amplitudes of two {\it sine} functions are not equal, the
results are more complicated but the basic conclusion about amplitude modulation remains.
When the power spectrum of $y(t)$ is constructed, it will show 3 peaks: the main peak at $f_o$ and
two more peaks at frequencies $f_o \pm \Delta f$. We used this fact to construct an algorithm for
automated searching for the evidence of amplitude  modulation. 
Fig \ref{fig:periodogram} compares the theoretical periodogram produced by interference beats with our algorithm's periodogram,
signifying that local Blazhko peaks are present in real data.

\begin{figure}[ht]
  \resizebox{\hsize}{!}{\includegraphics[width=17cm]{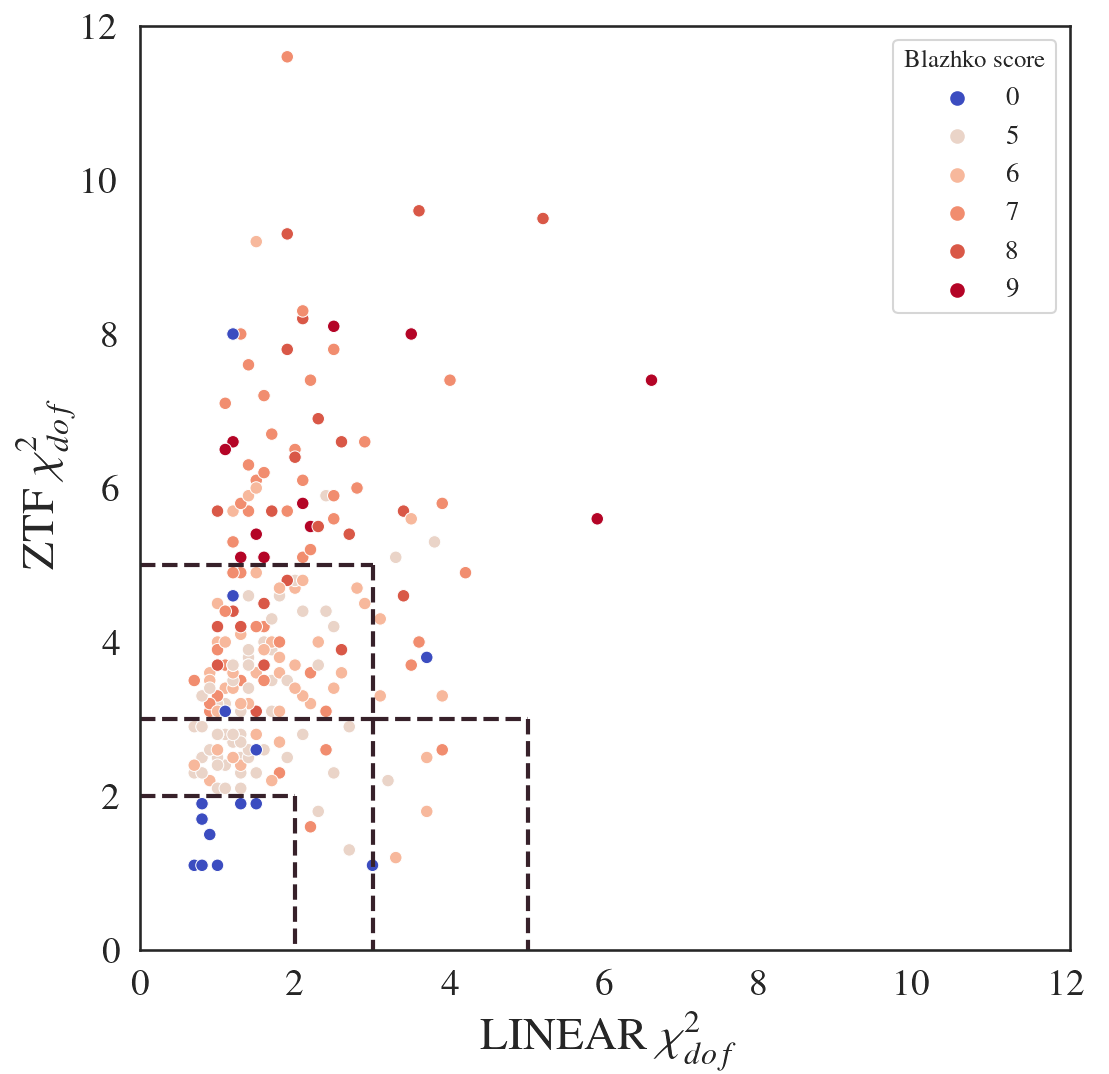}}
  \caption{Analogous to figure~\ref{fig:chi2}, except that here only
    228 visually verified Blazhko stars are shown.}
  \label{fig:chi_final}
  \end{figure}

After the strongest peak in the Lomb-Scargle periodogram is found at frequency $f_o$, we search for  two equally
distant local peaks at frequencies $f_-$ and $f_+$, with $f_- < f_0 < f_+$.  The sideband peaks can be highly asymmetric
\citep{2003ApJ...598..597A} and observed periodograms can sometimes be much more complex \citep{2007MNRAS.377.1263S}.  
We fold the periodogram through the main peak at $f_o$, multiply the two branches and then search for the strongest peaks
in the resulting folded periodogram that is statistically more significant than the background noise. The multiplication of the two
branches allows for efficient detection of moderately asymmetric peaks. However, this method is not sensitive to the subset of
modulated stars where only one sidepeak is detectable \citep{2018MNRAS.480.1229N, 2023A&A...678A.104M}.  The background noise
is computed as the scatter of the folded periodogram estimated from the interquartile range. We require a ``signal-to-noise''
ratio of at least 5, as well as the peak strength of at least 0.05 for ZTF, while 0.10 for LINEAR data.  If such a peak is found,
and it doesn't correspond to yearly alias, we select the star as a candidate Blazhko star and compute
its Blazhko period as 
\begin{equation*}
P_{BL} = |f_{-,+} - f_0|^{-1},
\end{equation*}
where $f_{-,+}$ means the Blazhko sideband frequency with a higher amplitude is chosen. 

The observed Blazhko periods range from 3 to 3,000 days, and Blazhko amplitudes range from 0.01 mag to about 0.3 mag \citep{2007MNRAS.377.1263S}. In this work, we selected a smaller Blazhko range due to the limitations of our data: 30--325 days. 
With this additional constraint, we selected 52 candidate Blazhko stars. 
Fig.~\ref{fig:periodogram} shows an example where two very prominent peaks were identified in the LINEAR periodogram.

\subsubsection{Visual Confirmation}

\begin{figure*}[ht]
  \centering
  \resizebox{\hsize}{!}{\includegraphics[width=17cm]{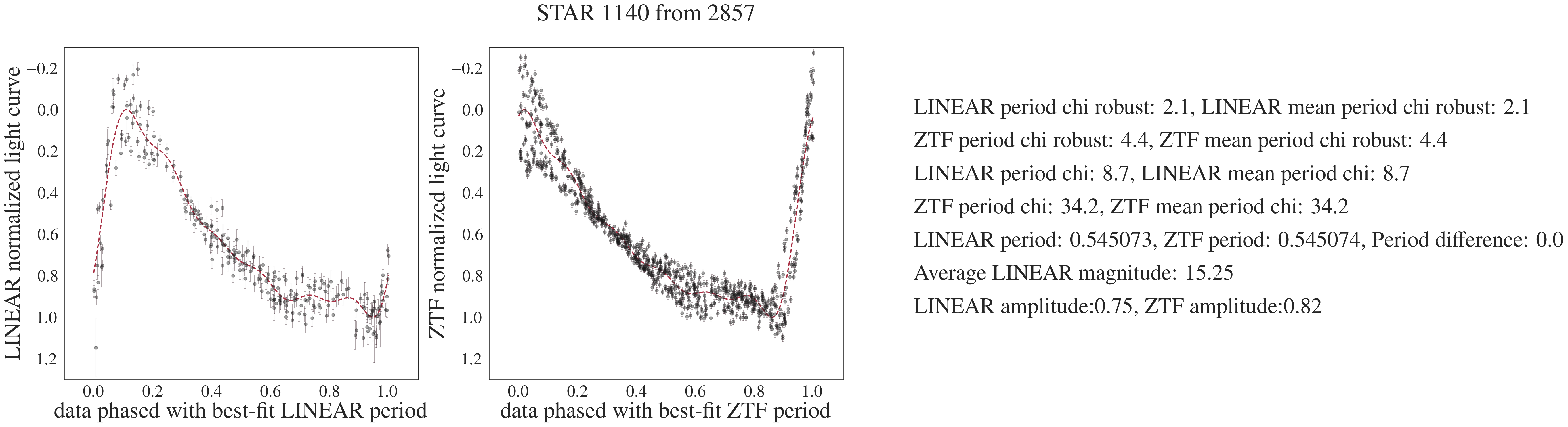}}
  \caption{An illustration of visual analysis of phased light curves for the selected Blazhko candidates. The left
    panel shows LINEAR data and the right panel shows ZTF data
    (symbols with ``error bars'') for star with LINEARid = 10030349. The dashed
    lines are best-fit models. The numbers listed on the right side were added to aid  visual analysis. Note
    multiple coherent data point sequences offset from the best-fit mean model in the right panel.}
       \label{fig:phase1}
\end{figure*}

\begin{figure*}[ht] 
    \centering
      \resizebox{\hsize}{!}{\includegraphics[width=17cm]{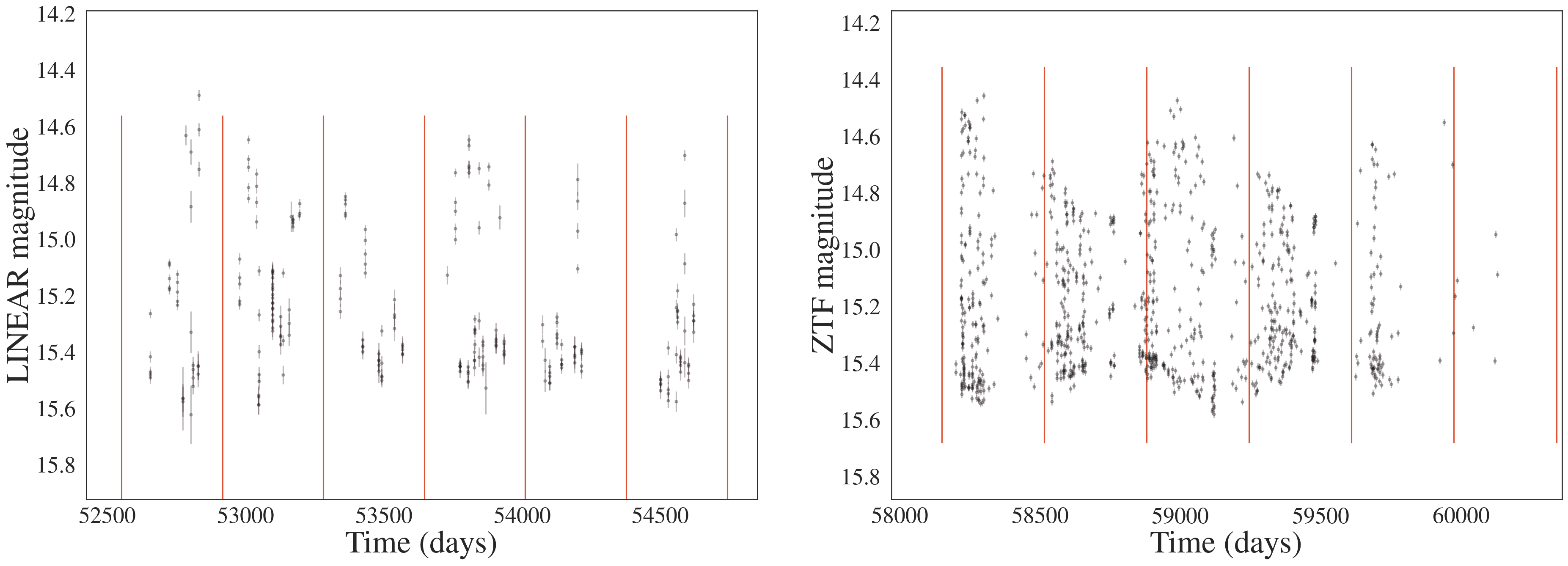}}
       \caption{An illustration of visual analysis of full light curves for the selected Blazhko candidates with emphasis
         on their repeatability between observing seasons, marked with  vertical lines (left: LINEAR data; right: ZTF data). Data
         shown are for star with LINEARid = 10030349. Note strong
         amplitude modulation between observing seasons.}
         \label{fig:phase3}
\end{figure*}
       
\begin{figure*}[ht]
    \centering
    \resizebox{\hsize}{!}{\includegraphics[width=16cm]{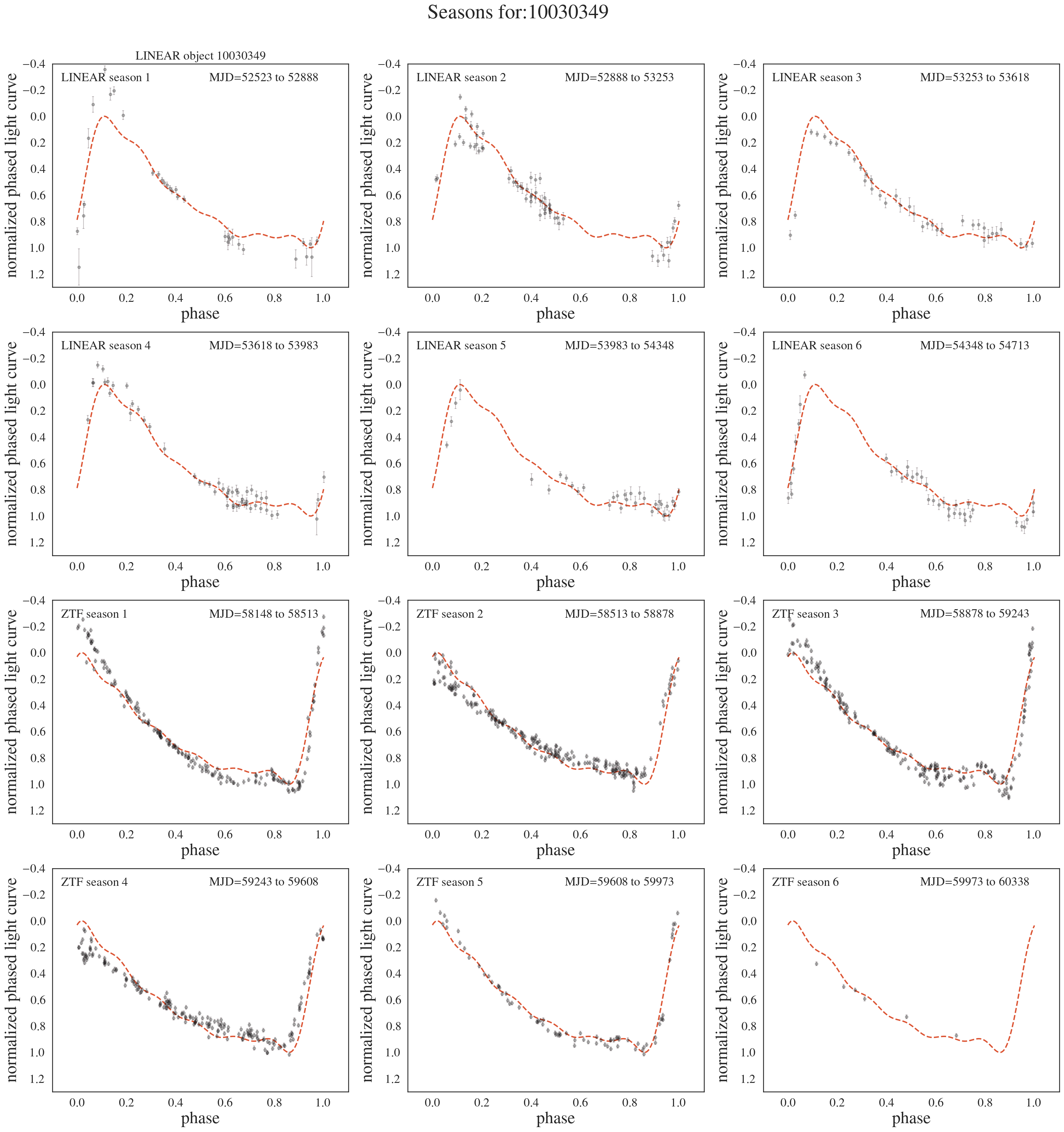}}
    \caption{The phased light curves normalized to unit amplitude of
      the overall best-fit model are shown for single observing seasons
      and compared to the mean best-fit models (top six panels: LINEAR data; bottom six panels: ZTF data).
      Data shown are for star with LINEARid = 10030349 (period = 0.54073 day). 
      Season-to-season phase and amplitude modulations are seen in both the LINEAR and the ZTF data.}
      \label{fig:phase4}
\end{figure*}

\begin{figure*}[ht]
    \centering
    \resizebox{\hsize}{!}{\includegraphics[width=16cm]{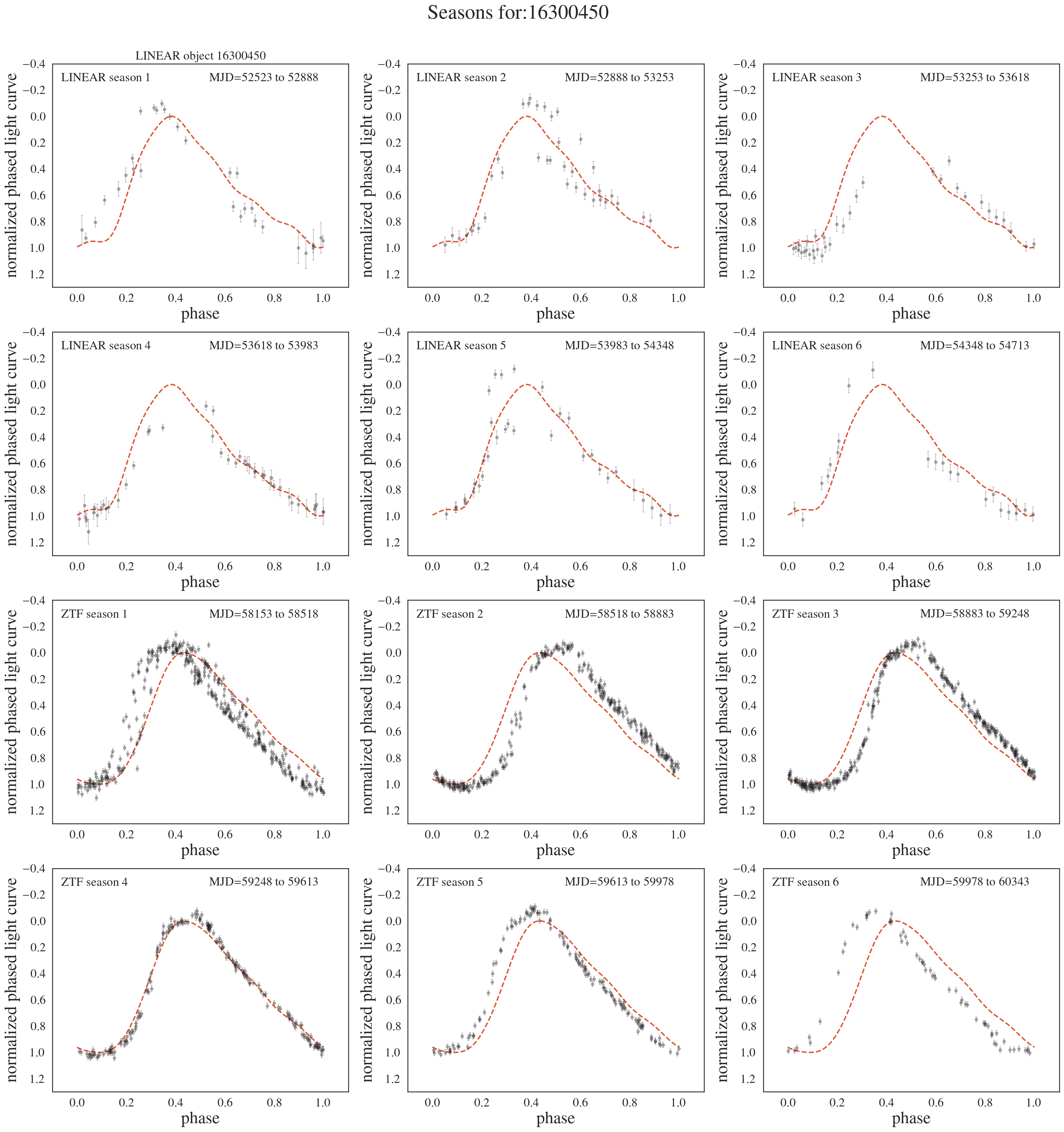}}
    \caption{Analogous to Fig.~\ref{fig:phase4}, except that star
    with LINEARid = 16300450 is shown (period = 0.33562 day). Unlike example shown in
    Fig.~\ref{fig:phase4},  only phase modulation is visible here,
    without any amplitude modulation, in both LINEAR and ZTF light curves.}
      \label{fig:phase5}
\end{figure*}

The sample pre-selected for visual analysis includes 531 RR Lyrae stars (479 + 52),
or 18.1\% of the starting LINEAR-ZTF sample. Visual analysis included the following standard steps
\citep[e.g.,][]{2009MNRAS.400.1006J, 2017MNRAS.466.2602P}: 
\begin{enumerate}
\item The shape of the phased light curves and scatter of data points around the best-fit model were examined
    for signatures of anomalous behavior indicative of the Blazhko effect. 
    Fig.~\ref{fig:phase1} shows an example of such behavior where the ZTF data and fit show multiple coherent data point sequences
    offset from the best-fit mean model. 
  \item Full light curves were inspected for their repeatability between observing seasons (Fig.~\ref{fig:phase3}).
       This step was sensitive to amplitude modulations with periods of the order a year or longer.  
     \item The phased light curves normalized to unit amplitude were inspected for their repeatability between observing seasons.
       This step was sensitive to phase modulations of a few percent or larger on time scales of the order a year or longer.  
       Fig.~\ref{fig:phase4} shows an example of a Blazhko star where season-to-season phase and amplitude modulations
       are seen in both the LINEAR data and (especially) the ZTF data. Another example is shown in Fig.~\ref{fig:phase5}
       where only phase modulation is visible,  without any discernible amplitude modulation\footnote{The physical reason
         for large phase modulations remains unclear: stellar companions may cause small variations through the
         light-time effect \citep{2021ApJ...915...50H}, but some RRc stars show phase variations well exceeding that
         \cite[e.g.,][]{2004MNRAS.354..821D, 2017MNRAS.465L...1S, 2019arXiv190200905L}.}. 
\end{enumerate}

After visually analyzing the starting sample of 531 Blazhko candidates, we visually confirmed expected Blazhko
behavior for 228 stars (214 out of 479 and 14 out of 52). LINEAR IDs and other characteristics for confirmed
Blazhko stars are listed in Table 1 (Appendix A). Statistical properties of the selected sample of Blazhko stars are
discussed in detail in the next section.

\section{Results}\label{sec:results}

\begin{figure*}[ht]
  \centering
  \resizebox{\hsize}{!}{\includegraphics[width=16cm]{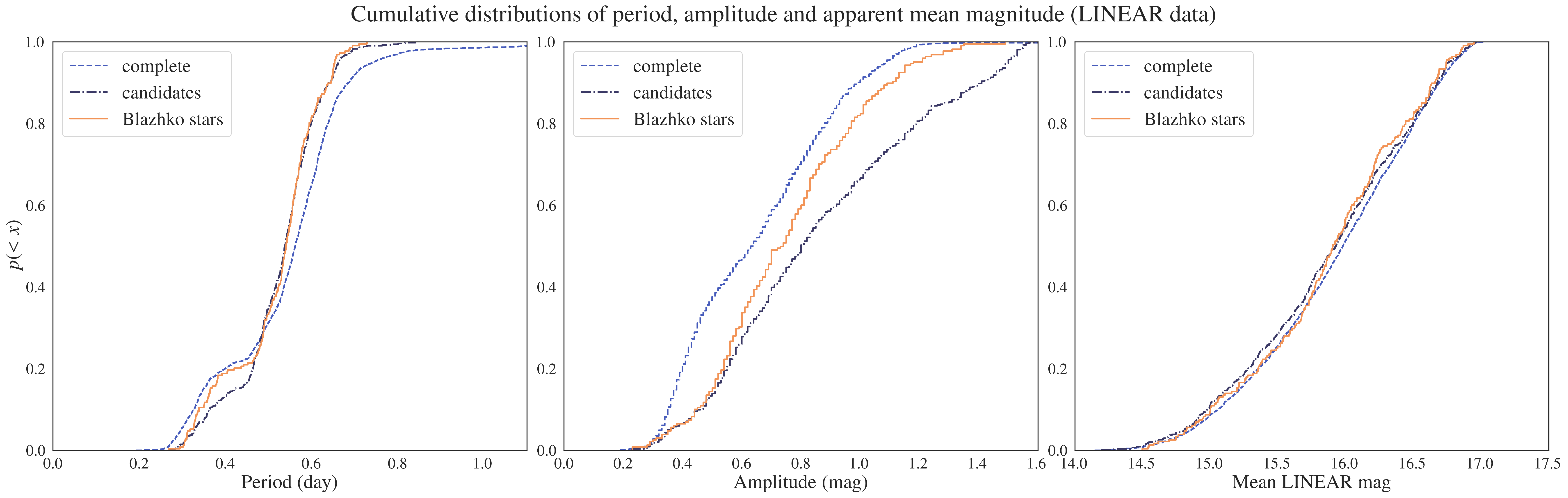}}
   \resizebox{\hsize}{!}{\includegraphics[width=16cm]{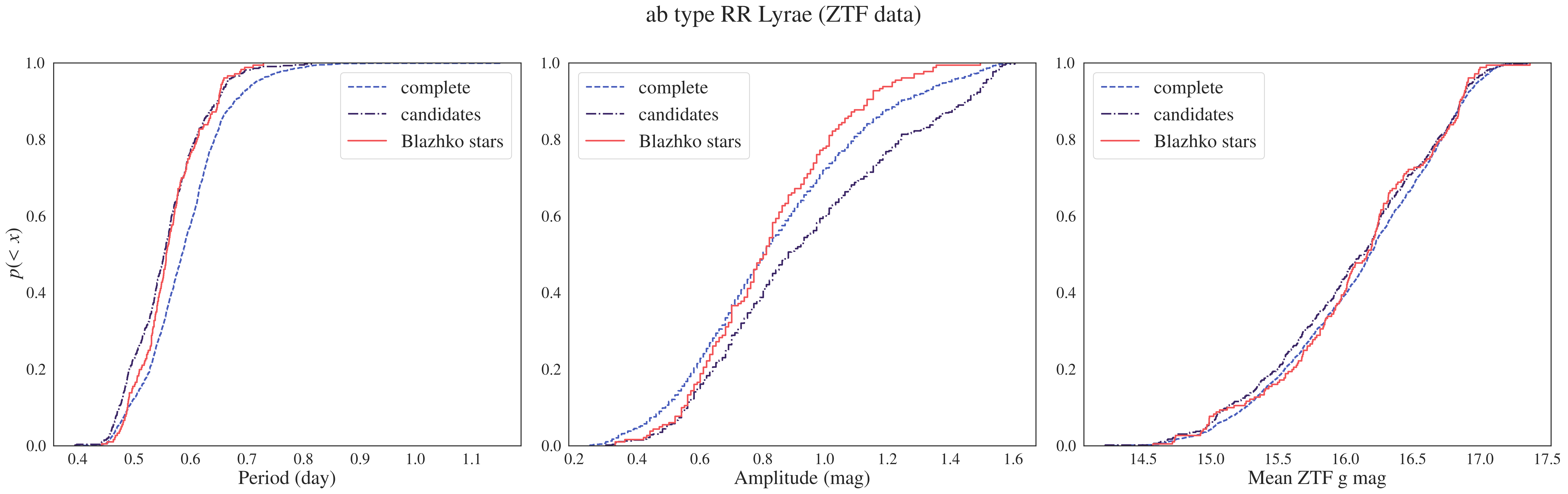}}
   \resizebox{\hsize}{!}{\includegraphics[width=16cm]{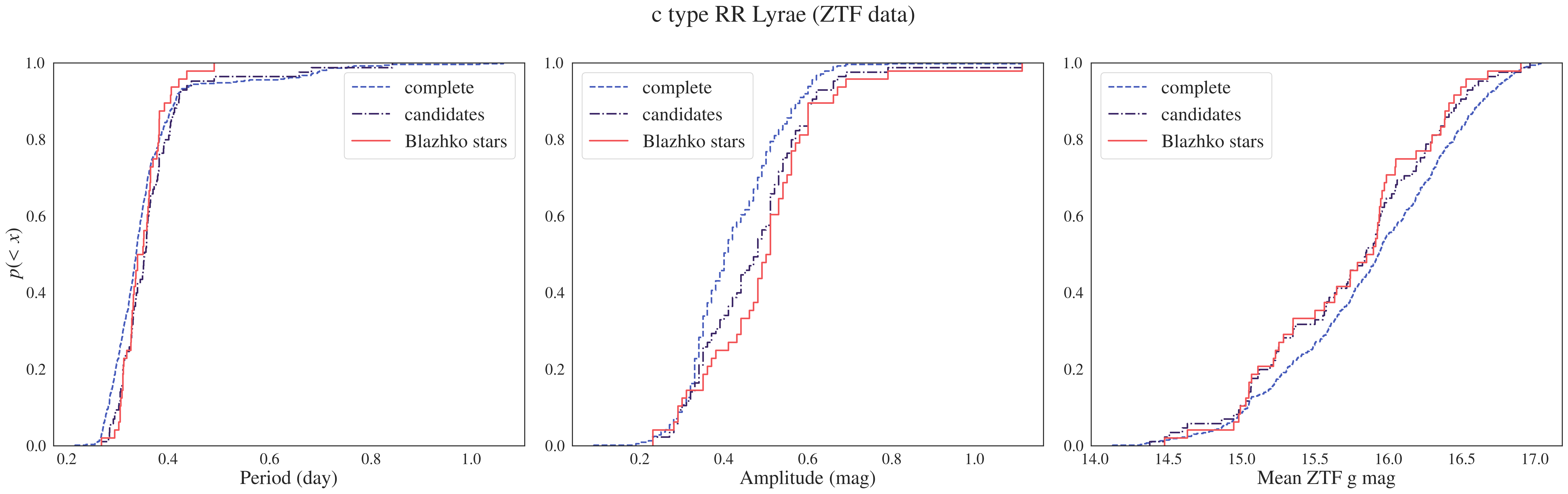}}
 \caption{A comparison of cumulative distributions of period (left),
 amplitude (middle) and apparent magnitude for starting sample,
 selected Blazhko candidates and visually verified Blazhko
 stars. The top row is based on LINEAR data and both ab type and c
 type stars. The middle and bottom rows are
 based on ZTF data, and show separately data for ab type and c type
 stars, respectively. The differences in period and amplitude
 distributions are futher examined in figure~\ref{fig:AmplPeriod2D}.}
    \label{fig:AmplPeriod}
\end{figure*}

\begin{figure*}[ht]
  \centering
  \resizebox{\hsize}{!}{\includegraphics[width=16cm]{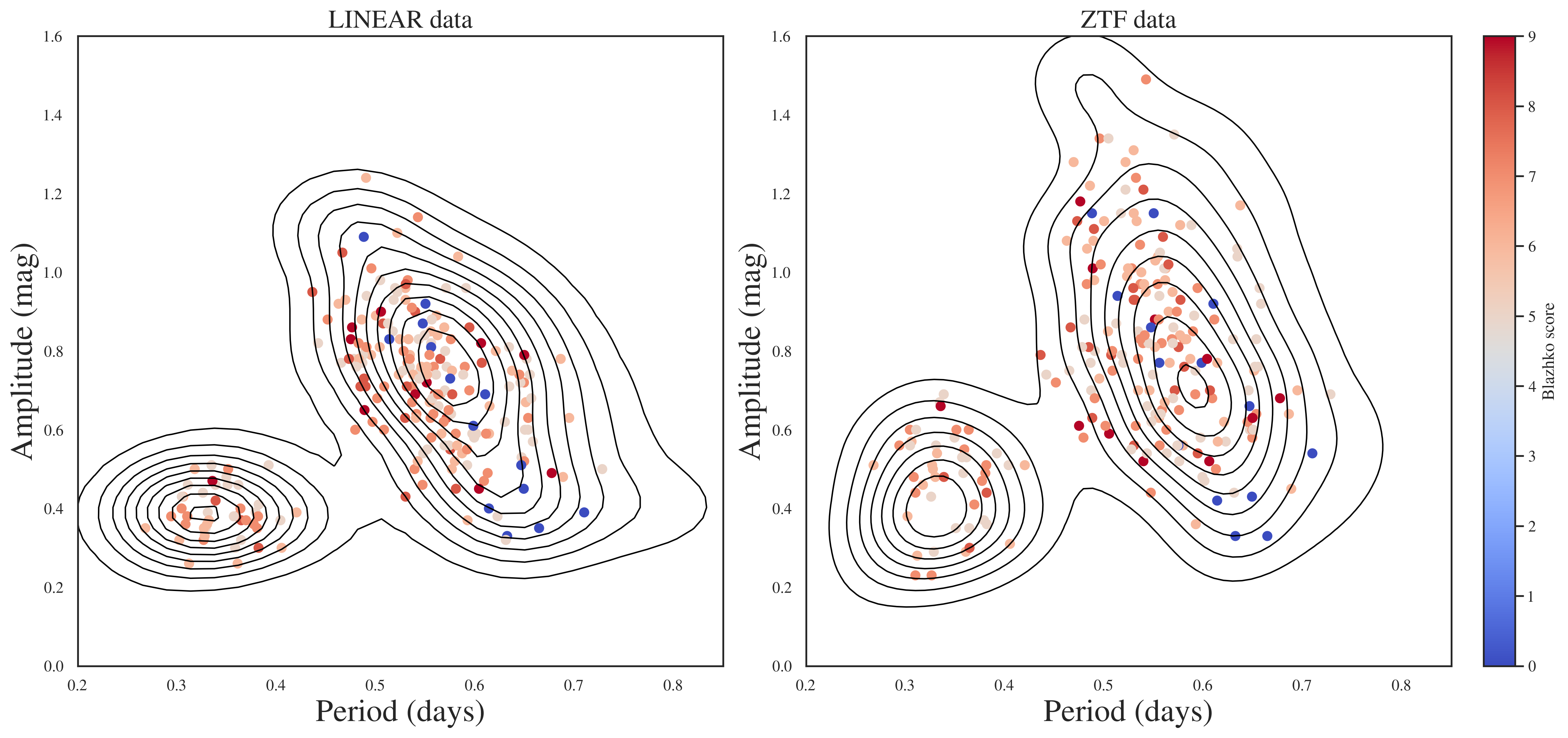}}
  \caption{Comparison of amplitude--period distributions (the Bailey
    diagram) for the starting sample of 1,996 RR Lyrae stars (contours)
      and 228 selected candidate Blazhko stars (symbols). The clump
      in the lower left corresponds to c type RR Lyrae and the
      other one to ab type. Note that the period distribution for ab
    type Blazhko stars is shifted left (by about 0.03 day, or 5\%).}
    \label{fig:AmplPeriod2D}
\end{figure*}

\begin{figure*}[ht]
  \centering
  \resizebox{\hsize}{!}{\includegraphics[width=16cm]{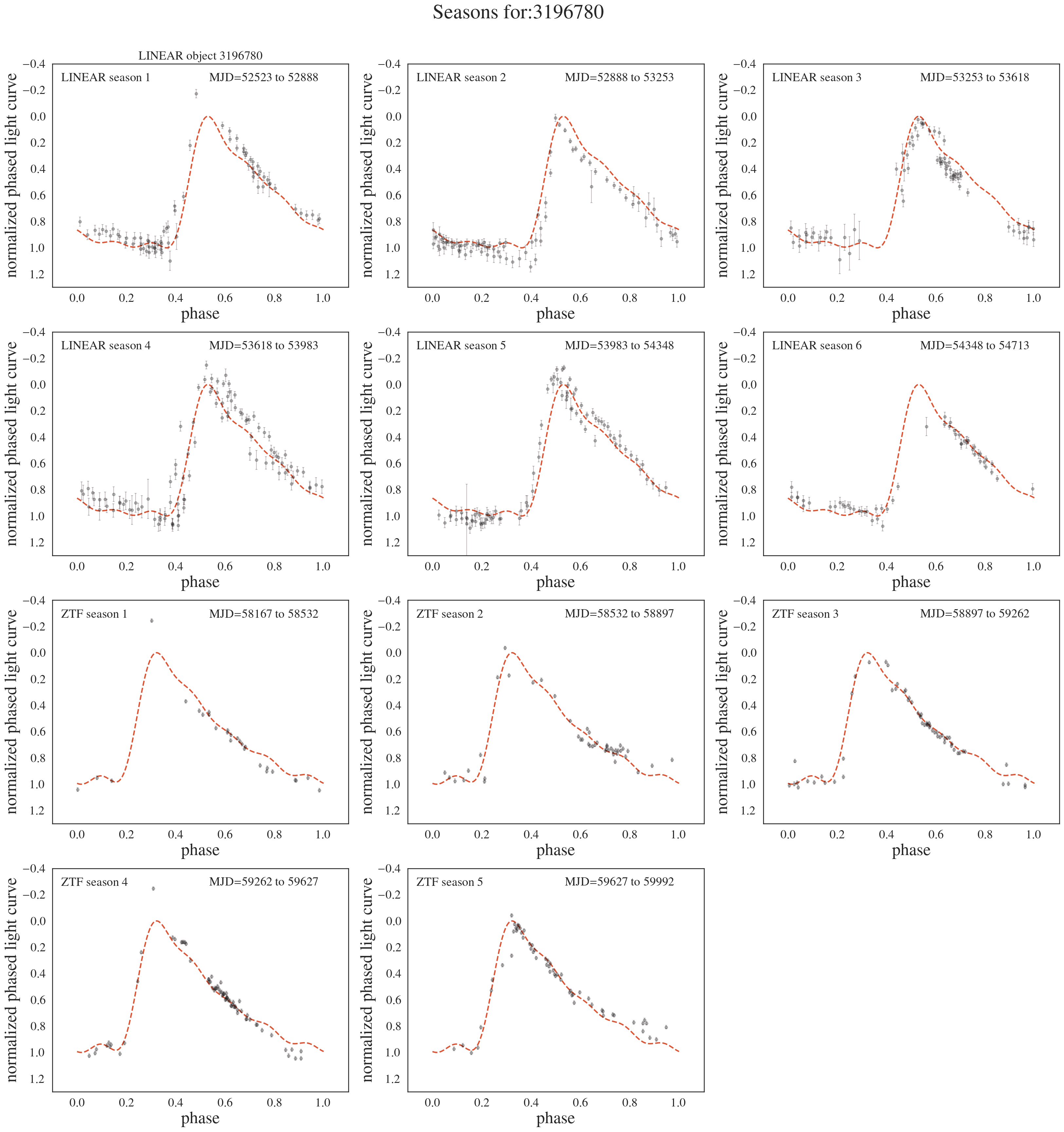}}
  \caption{Analogous to Fig.~\ref{fig:phase4}, except that star
  with LINEARid = 3196780 is shown. Amplitude modulation is clearly
  seen in LINEAR light curves (top two rows), while not discernible
  in ZTF light curves (bottom two rows). Additional stars with
  similar behavior include LINEARid = 2889542, 7723614, 8342007.
  This behavior strongly  suggests that Blazhko effect can appear and disappear on time scales shorter than
    about a decade.
    }
    \label{fig:phase6}
\end{figure*}

Starting with a sample of 2857 field RR Lyrae stars with both LINEAR and ZTF data, we constructed a subsample of
1996 with light curves of sufficient quality and selected and verified 228 stars that exhibit convincing Blazhko effect.
In this section we compare various statistical properties of selected Blazhko stars to those of the starting sample. 

\subsection{The Blazhko Incidence Rate}

The implied incidence rate for the Blazhko effect is
11.4$\pm$0.8\%. Due to selection effects and unknown completeness,
this rate should be considered as a lower limit. 
When ab and c types are considered separately, the
rate is slightly higher for the former than for the latter: 12.1$\pm$0.9\%
vs. 9.2$\pm$1.3\%.  The difference of 2.9\% has low statistical significance ($<2\sigma$).

\subsection{Period, Amplitude and Magnitude Distributions}

Marginal distributions of  period, amplitude and apparent magnitude
for the starting sample and Blazhko stars are compared in Fig.~\ref{fig:AmplPeriod}. 
Encouragingly, their magnitude distributions are statistically
indistinguishable which indicates that the completeness is not a
strong function of the photometric signal-to-noise ratio. This
result is probably due to the fact that the sample is defined by the
depth of LINEAR survey, while ZTF survey is deeper than this limit and
its photometric quality is approximately constant across the probed
magnitude range. 

The suspected differences in amplitude and period distributions are
further explored in Fig.~\ref{fig:AmplPeriod2D}. It is already
discernible by eye that the period distribution for Blazhko stars of
ab type is shifted to smaller values than for the starting sample. We have
found that the median period for ab type Blazhko stars is about 5\% shorter
than for the starting RR Lyrae sample. This difference is significant at
the 7.1$\sigma$ level.  At the same time, the difference in median
amplitudes for ab type stars corresponds to only 0.6$\sigma$ deviation. 
No statistically significant differences are found in period and
amplitude distributions for c type stars.

If modulation amplitudes are correlated with periods such that larger
modulation amplitudes occur in shorter period RRab stars, and if
our selection efficiency is lower for smaller modulation amplitudes,
then the detected period shift for ab type Blazhko stars might be at
least partially due to combination of these two effects. This
possibility does not appear likely. First, as we discussed in
preceeding section, our sample is defined by the depth of LINEAR survey, 
while ZTF survey is significantly deeper than this limit and its photometric quality is approximately constant across the probed 
magnitude range. Since it is sufficient for a star to display the
Blazhko effect only in ZTF to be included in the sample, 
we do not expect strong selection effects (except for the LINEAR
magnitude cutoff of course).  Furthermore, \cite{2020MNRAS.494.1237S}
searched for period - modulation amplitude correlation using a large
sample of stars with OGLE measurements and did not find any.

\subsection{Long-term behavior of Blazhko Stars}

During visual analysis, we noticed that some Blazhko stars exhibit
convincing Blazhko effect either in LINEAR or in ZTF data, but not in
both surveys. Fig.~\ref{fig:phase6} shows an example where amplitude
modulation is clearly seen in LINEAR light curves, while not discernible
in ZTF light curves.  There are also examples of stars where Blazhko effect is evident in
ZTF but not in LINEAR data (e.g., LINEARid = 19466437, 14155360). 
This finding  strongly suggests that Blazhko effect can appear and disappear on time scales shorter than
about a decade.

\section{Discussion and Conclusions\label{sec:discussion}}

We found excellent agreement
between the best-fit periods for RR Lyrae stars estimated separately from LINEAR and ZTF light curves. 
Only one star in our sample (CT CrB, LINEARid=17919686), was previously reported as a Blazhko star \citep{2013A&A...549A.101S}.
The sample of 228 stars presented here increases the number of field RR Lyrae stars displaying the Blazhko
effect by more than 50\% and places a lower limit of (11.4$\pm$0.8)\%
for their incidence rate. The reported incidence rates for the Blazhko effect
range from 5\% \citep{2007MNRAS.377.1263S} to 60\% \citep{2014A&A...570A.100S}.
Differences in reported incidence rates can occur due to varying data precision, the temporal baseline length, and differences in visual or algorithmic analysis.
For a relatively small sample of
151 stars with Kepler data, a claim has been made that essentially every RR Lyrae star exhibits modulated light curve
\citep{2018A&A...614L...4K}. The difference in Blazhko incidence rates for the two most extensive samples, obtained
by the OGLE-III survey for the Large Magellanic Cloud (LMC, 20\% out of 17,693 stars; \citealt{2009AcA....59....1S})
and the Galactic bulge (30\% out of 11,756 stars; \citealt{2011AcA....61....1S}) indicates a possible variation of
the Blazhko incidence rate with underlying stellar population properties. 
 
We find that ab type RR Lyrae which show the Blazhko effect have about 5\% (0.030 day) shorter periods than starting
sample. While not large, the statistical significance of this difference is 7.1$\sigma$. At a similar uncertainty level
($\sim$1\%), we don't detect period difference for c type stars, and don't detect any difference in amplitude distributions.
We also find that for some stars the Blazhko effect is discernible in only one dataset. This finding  strongly suggests that Blazhko effect can
appear and disappear on time scales shorter than about a decade, in agreement with literature 
\citep{2009MNRAS.400.1006J, 2010A&A...520A.108P, 2014ApJS..213...31B}.

The LINEAR and ZTF datasets analyzed in this work were sufficiently large that we had to rely on algorithmic
pruning of the initial sample. The sample size problem will be even larger for surveys such as the Legacy Survey
of Space and Time (LSST; \citealt{2019ApJ...873..111I}). LSST will be an excellent survey for studying Blazhko effect
\citep{2022ApJS..258....4H} because it will have both a long temporal baseline (10 years) and a large number of
observations per object (nominally 825; LSST Science Requirements Document\footnote{Available as ls.st/srd}).
We anticipate a higher fraction of discovered Blazhko stars with LSST than reported here due to better sampling
and superior photometric quality, since the incidence rate of the Blazhko effect increases with sensitivity to
small-amplitude modulation, and thus with photometric data quality \citep{2009MNRAS.400.1006J}.

The size and quality of LSST sample will motivate further developments of the selection algorithms. 
One obvious improvement will be inspection of neighboring objects to confirm photometric quality,
as well as inspection of images to test implication of an isolated point source (e.g., blended object photometry
can be affected by variable seeing beyond aperture correction valid for isolated point sources). 
Another improvement is forward modeling of the Blazhko modulation, rather than searching for $\chi^2$
outliers \citep{2011MNRAS.417..974B, 2012MNRAS.424..649G}.
For example, \cite{2020MNRAS.494.1237S} classified Blazhko stars in 6 classes using the morphology
of their amplitude modulation (the most dominant class includes 90\% of the sample). They also found bimodal distribution
of Blazko periods, with two components centered on 48 d and 186 d. These results give hope that forward
modeling of the Blazhko effect will improve the selection of such stars.

%\begin{acknowledgements}
\section*{Acknowledgments} 
We thank Mathew Graham for providing {\it ztfquery} code example to
us, and Robert Szab{\'o} for expert comments that improved presentation. 

\v{Z}.I. acknowledges funding by the Fulbright Foundation and thanks the Ru\d er Bo\v{s}kovi\'{c} Institute (Zagreb, Croatia) for hospitality.

This material is based on work supported in part by the
National Science Foundation through Cooperative Agreement
1258333 managed by the Association of Universities for
Research in Astronomy (AURA), and the Department of
Energy under Contract No. DE-AC02-76SF00515 with the
SLAC National Accelerator Laboratory. Additional LSST
funding comes from private donations, grants to universities,
and in-kind support from LSSTC Institutional Members. This
research has made use of NASA’s Astrophysics Data System
Bibliographic Services.

Based on observations obtained with the Samuel Oschin Telescope 48-inch and the 60-inch Telescope at the Palomar Observatory as part of the Zwicky Transient Facility project. ZTF is supported by the National Science Foundation under Grants No. AST-1440341 and AST-2034437 and a 
collaboration including current partners Caltech, IPAC, the Weizmann Institute of Science, the Oskar Klein Center at Stockholm University, the University of Maryland, Deutsches Elektronen-Synchrotron and Humboldt University, the TANGO Consortium of Taiwan, the University of Wisconsin at Milwaukee, Trinity College Dublin, Lawrence Livermore National Laboratories, IN2P3, University of Warwick, Ruhr University Bochum, Northwestern University and former partners the University of Washington, Los Alamos National Laboratories, and Lawrence Berkeley National Laboratories. Operations are conducted by COO, IPAC, and UW.

The LINEAR program is funded by the National Aeronautics and Space Administration at MIT Lincoln Laboratory under Air Force Contract FA8721-05-C-0002.
Opinions, interpretations, conclusions and recommendations are those of the authors and are not necessarily endorsed by the United States Government.
%\end{acknowledgements}

\software{Astropy \citep{2018AJ....156..123A, astropy:2022},
         Matplotlib \citep{matplotlib2007},
         SciPy \citep{scipy2020}, 
         astroML \citep{2012cidu.conf...47V}}

%\vspace{5mm}
%\facilities{LINEAR, Zwicky Transient Factory}
%\software{Astropy \citep{astropy:2013, astropy:2018, astropy:2022}, Matplotlib \citep{matplotlib2007}, SciPy \citep{scipy2020} }

\appendix
Table 1: The first 10 confirmed Blazhko stars with their LINEAR
IDs in the first column and then, for both LINEAR and ZTF, their
computed  light curve periods (day),
the number of data points per light curve, robust and ordinary $\chi^2$ values, and light curve amplitudes, followed by amplitude 
difference between LINEAR and ZTF, the strength and period of Blazhko peaks in their periodograms, light curve type (1: ab, 2: c), detection
significance flag for the periodograms (Z, L or ``-'' for no detection; the strength and period of Blazhko
peaks are not reliable when ``-'' ) and the selection score (see Sections 3.1 and 3.2 for details). The full table is available in online edition.

%\input{be_full.tex}
% only the first 10 entries from be_full.tex
\begin{center}
%\footnotesize
\scriptsize
\begin{tabular}{rccrrccrrrrrrrrrrlrr}
\toprule
 LID &  P$_L$ &   P$_Z$ &  N$_L$ &  N$_Z$&  $\chi^2_{L,r}$ &  $\chi^2_{Z,r}$ &  $\chi^2_L$ &  $\chi^2_Z$ & $A_L$ &  $A_Z$ &  $\delta$A &  Bp$_L$ &  Bp$_Z$ &  Bp$_L$ &  Bp$_Z$ &  t &  f &  B\_s & B\_f \\
\midrule
    158779 & 0.609207 & 0.609189 &  293 &  616 &      1.6 &      3.9 &     3.7 &    34.2 &   0.47 &   0.68 &       0.21 &  1.6443 &  1.6444 &  352.7337 &  350.2 &       1 &             - &        7 &         1 \\
    263541 & 0.558218 & 0.558221 &  270 &  503 &      2.9 &      6.6 &    15.8 &   110.4 &   0.64 &   0.82 &       0.18 &  1.8621 &  1.8025 &   14.1513 &   89.9 &       1 &             - &        7 &         1 \\
    393084 & 0.530027 & 0.530033 &  493 &  372 &      1.1 &      3.2 &     1.6 &    19.2 &   0.96 &   1.31 &       0.35 &  1.9447 &  1.8896 &   17.2369 &  347.2 &       1 &             - &        6 &         1 \\
    810169 & 0.465185 & 0.465212 &  289 &  743 &      2.1 &      2.8 &     6.0 &    15.1 &   0.77 &   0.75 &       0.02 &  2.2232 &  2.2230 &   13.6017 &   13.6 &       1 &             - &        5 &         1 \\
    924301 & 0.507503 & 0.507440 &  418 &  189 &      1.9 &      9.3 &    13.8 &   162.9 &   0.87 &   0.79 &       0.08 &  2.0043 &  1.9763 &   29.5072 &  178.4 &       1 &             - &        8 &         1 \\
    970326 & 0.592233 & 0.592231 &  275 &  552 &      1.1 &      2.1 &     1.9 &     7.7 &   0.51 &   0.75 &       0.24 &  1.7563 &  1.6992 &   14.7656 &   93.2 &       1 &             - &        5 &         1 \\
    999528 & 0.658401 & 0.658407 &  564 &  213 &      1.2 &      2.7 &     1.8 &    21.7 &   0.57 &   0.92 &       0.35 &  1.5527 &  1.5510 &   29.5247 &   31.0 &       1 &             - &        5 &         1 \\
   1005497 & 0.653607 & 0.653605 &  607 &  192 &      1.1 &      2.1 &     2.1 &    12.4 &   0.60 &   0.83 &       0.23 &  1.5639 &  1.5481 &   29.4638 &   55.1 &       1 &             - &        5 &         1 \\
   1092244 & 0.649496 & 0.649558 &  590 &  326 &      1.2 &      3.6 &     2.3 &    32.1 &   0.72 &   0.58 &       0.14 &  1.5735 &  1.5640 &   29.5421 &   40.8 &       1 &             - &        7 &         1 \\
   1240665 & 0.632528 & 0.632522 &  468 &  311 &      3.0 &      1.1 &    25.2 &     1.6 &   0.33 &   0.33 &       0.00 &  1.6149 &  1.5865 &   29.4942 &  182.3 &       1 &             Z &        0 &         2 \\
\bottomrule
\end{tabular}
\end{center}

\bibliographystyle{aasjournal} % style aa.bst
\bibliography{paper} % your references Yourfile.bib
\end{document}